%% file: main.tex
\pdfoutput=1
\documentclass[11pt]{article}

\usepackage{acl}

\usepackage{times}
\usepackage{placeins} 
\usepackage{latexsym}
\usepackage{amsmath}
\usepackage{amssymb}
\usepackage{mathtools}
\usepackage{amsthm}

\usepackage{algorithm}
\usepackage{algorithmic}
\usepackage{float} 
\usepackage{subcaption}  % Allows subfigures

\usepackage[T1]{fontenc}
\usepackage[utf8]{inputenc}
\usepackage{microtype}

\usepackage{inconsolata}

\usepackage{graphicx}
\usepackage[textsize=tiny]{todonotes}

\title{CAARMA: Class Augmentation with Adversarial Mixup Regularization}

\author{
  Massa Baali, Xiang Li, Hao Chen, Syed Abdul Hannan, Rita Singh, Bhiksha Raj \\
  Carnegie Mellon University \\
  \texttt{mbaali@cs.cmu.edu}
}

\begin{document}
\maketitle

\begin{abstract}

\input{sections/abstract}
\end{abstract}

\section{Introduction}\label{sec:intro}
\input{sections/intro_v2}

\section{Related-Work}
% [aDD THIS: generated fake speakers to build sv ssytems]
\input{sections/related-work}

\section{Class Augmentation with Adversarial Mixup Regularization}
\input{sections/method_2}

\section{Experiments}
\input{sections/exp2}

\section{Results \& Analysis}
\input{sections/results}

\section{Conclusion}
\input{sections/conclusion}
\section*{Limitations}
The CAARMA framework, while showcasing notable enhancements in speaker verification and zero-shot learning tasks, is subject to several limitations that merit further exploration. Although it  performs well in controlled settings, its scalability to extremely large or diverse datasets, as well as its applicability to real-world scenarios with high speaker variability, has yet to be fully established. This also adds complexity to the implementation and increases computational demands, which may restrict accessibility for those with limited resources. 

\section*{Ethics Statement}
The CAARMA framework is developed with a commitment to ethical considerations, especially concerning privacy and the potential for surveillance misuse. It is crucial to ensure that this technology, while advancing the capabilities of speaker verification systems, is employed within the confines of strict ethical guidelines and privacy regulations to prevent any invasion of individual privacy. As this framework facilitates the generation of synthetic data, we also focus on preventing biases that could arise in synthetic datasets, ensuring fair representation across different groups. In adherence to the ACL Ethics Policy, we emphasize transparency in the deployment of CAARMA and advocate for its use in ethically justifiable manners that respect individual rights and data integrity.

\newpage
\bibliography{main}

\appendix

% \section{Example Appendix}
% \label{sec:appendix}

% This is an appendix.

\end{document}

%% file: sections/abstract.tex
Speaker verification is a typical zero-shot learning task, where inference of unseen classes is performed by comparing embeddings of test instances to known examples. The models performing inference must hence naturally generate embeddings that cluster same-class instances compactly,  while maintaining separation across classes. In order to learn to do so, they are typically trained on a large number of classes (speakers), often using specialized losses.  However real-world speaker datasets often lack the class diversity needed to effectively learn this in a generalizable manner. We introduce CAARMA,  a class augmentation framework that addresses this problem by generating synthetic classes through data mixing in the embedding space, expanding the number of training classes. To ensure the authenticity of the synthetic classes we adopt a novel adversarial refinement mechanism that minimizes categorical distinctions between synthetic and real classes. We evaluate CAARMA on multiple speaker verification tasks, as well as other representative zero-shot comparison-based speech analysis tasks and obtain consistent improvements: our framework demonstrates a significant improvement of 8\% over all baseline models. The code is available at: \url{https://github.com/massabaali7/CAARMA/} % The code for CAARMA will be open-sourced.
% and made available for public use.
%These results highlight the potential of leveraging synthetic data and manifold regularization to enhance zero-shot learning performance and adaptability in machine learning systems.

%% file: sections/intro_v2.tex
 %%%The challenge of training models for effective zero-shot learning (ZSL) lies in generating embeddings that cluster same-class instances while maintaining separation between different classes \cite{Zhu_2019_CVPR}. To achieve this, they are typically trained on a large number of classes, often using specialized losses. The underlying principle here is that by learning from a sufficiently large number of classes, and through proper encouragement embodied in the losses, the model learns not merely to separate the classes it \textit{has} seen, but the more general principle that instances from a class must be clustered closely together while begin separated from those from other classes [CITES]. 
 %
 %%%However, real-world training datasets often lack the necessary variety of classes, severely limiting the model’s ability to develop robust and transferable representations \cite{xie2022towards}.
% Speaker verification is, in essence, a zero-shot learning (ZSL) task --  verification is performed simply by matching embeddings derived from enrollment and verification instances with no additional training \cite{wan2018generalized}. Consequently, the discussion below is generically stated in terms of ZSL; it must nevertheless be understood that  our focus remains on the problem of speaker verification.

% We propose a novel augmentation-based training paradigm for Speaker Verification systems to improve training in low-diversity training data regimes.
Speaker verification is fundamentally a zero-shot learning (ZSL) task, where verification is accomplished by comparing embeddings from enrollment and verification samples without the need for further training \cite{wan2018generalized}. This process aligns with the principles of ZSL, where models are expected to operate effectively on unseen data. Therefore, while the following discussion is framed within the broader context of ZSL, it is specifically tailored to address the challenges in speaker verification.

To address the challenge of limited class diversity in training datasets, a common issue in speaker verification, we propose a novel augmentation-based training paradigm. This approach leverages synthetic data augmentation to enhance the robustness and generalization capabilities of speaker verification systems, particularly in low-diversity environments. Our method not only stays true to the essence of ZSL by facilitating effective generalization to new speakers but also introduces a practical solution to overcome the inherent limitations of traditional training datasets.
\begin{figure}[t]
\begin{center}
\centerline{\includegraphics[width=1.1\linewidth]{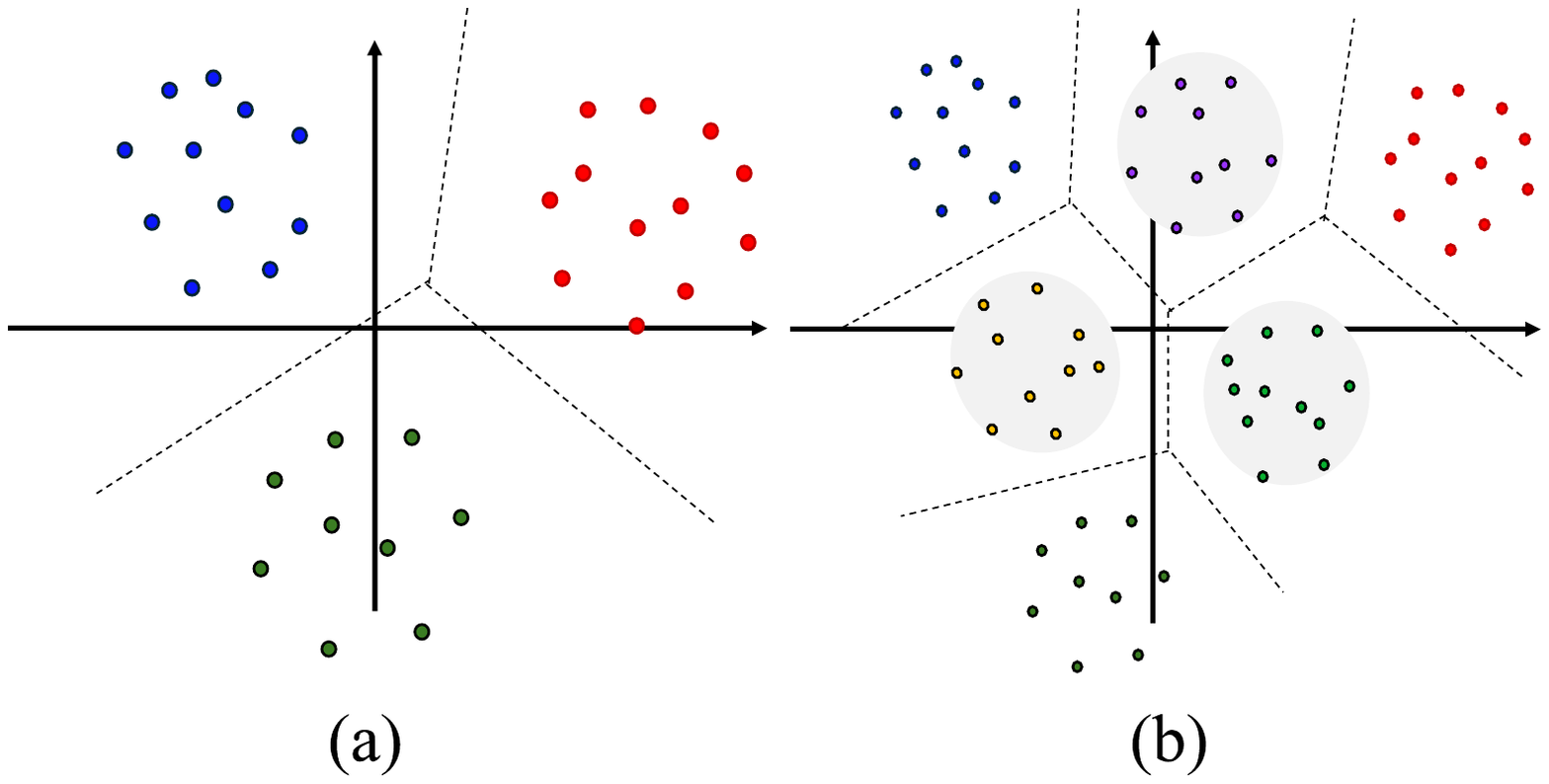}}
\vspace{-1.0cm}
\caption{(a) When trained with fewer classes the model can spread the embeddings of individual classes out while still learning to classify the training data accurately and with large margins. This will not, however translate to compact representations for newer unseen classes. (b) With additional synthetic classes (shaded grey), the model must now learn to compact classes more. This will translate to more compact unseen classes as well.}
\label{fig:main_figure}
\end{center}
\vspace{-1.0cm}
\end{figure}

Effective zero-shot learning lies in generating embeddings that cluster same-class (in our case, same-speaker) instances closely while maintaining separation between different classes \cite{Zhu_2019_CVPR}. Traditional training approaches for ZSL models rely on two key components: exposure to a large number of diverse classes and the use of specialized loss functions that promote both inter-class separation and intra-class compactness \cite{min2020domain}. The underlying principle here is that by learning from a sufficiently large number of classes, and through proper encouragement embodied in the losses, the model learns not merely to separate the classes it has seen, but \textit{the more general principle} that instances from a class must be clustered closely together while begin separated from those from other classes \cite{xian2018zero}. 

 However, when the training datasets lack the necessary variety of classes (speakers), this can severely limit the model’s ability to develop robust and transferable representations \cite{xie2022towards}. Indeed, it may be argued that in the high-dimensional space of the embeddings, \textit{any} finite set of training classes is insufficient to cover the space adequately.
This limitation leads to models that fail to generalize effectively to unseen categories, resulting in suboptimal zero-shot inference performance \cite{gupta2021generalized}.

Popular approaches to address training data limitations often rely on data augmentation techniques. Methods such as AutoAugment \cite{cubuk2019autoaugment} and SpecAugment \cite{park2019specaugment} generate new samples by modifying existing ones through transformations like geometric distortions, time warping, and frequency masking. However, while these techniques increase intra-class diversity, they do not introduce new classes, limiting their effectiveness in zero-shot learning scenarios.
Of most relevance to our paper are \textit{mixup}-based data augmentation techniques, \textit{e.g.} \cite{verma2019manifold, yun2019cutmix}, that aim to enhance training by generating new samples through interpolation \cite{han2021contrastive} of both the features and their labels. Regardless of the interpolation, yet these methods also do not generate \textit{new} classes;  they merely improve the generalization of the model by mapping mixed data to mixed-class labels.  Still other methods use generative models such as Variational Autoencoders (VAEs), Generative Adversarial Networks (GANs), diffusion models \textit{etc.} to generate authentically novel data to enhance the training \cite{ min2019domain}; however these too are generally restricted to generating novel instances for \textit{known} classes, limiting their effectiveness in zero-shot scenarios \cite{pourpanah2022review}. Thus, while these approaches are generally very successful in improving generalization in classification problems, they fail at addressing the problem ZSL learning faces, that of increasing the number of classes themselves,  leading to inconsistent generalization to unseen classes \cite{xie2022towards}. 

In this paper, we introduce \textit{Class Augmentation with AdversaRial Mixup regulariAztion} (CAARMA), a data augmentation framework to introduce \textit{synthetic} classes (speakers) to enhance ZSL training  for speaker verification.  CAARMA utilizes a mixup-like strategy to generate data from fictitious speakers.  However, unlike conventional mixup which mixes data in the input space, which would arguably be meaningless in our setting (a straight-forward mix of two speech recordings will merely result in a mixed signal, and not a new speaker), the mixup is performed in the embedding space in a manner that permits assignment of new class identities to the mixed-up data. Critically, we must now ensure that the mixed-up embeddings resemble those from actual speakers. We do so through a discriminator that is used to minimize categorical distinctions between synthetic and authentic data through adversarial training.

We demonstrate CAARMA's effectiveness through extensive evaluation on speaker verification, where it achieves substantial improvements in generalizing to diverse speaker distributions. Additional experiments on other ZSL speech tasks further validate our approach's broad applicability.

Our main contributions are as follows:
\begin{itemize}
\item We introduce CAARMA, a novel class augmentation framework that addresses the fundamental limitation of class diversity in zero-shot learning by generating synthetic classes termed as Sythetic Label Mixup (SL-Mixup) through embedding-space mixing, rather than conventional input-space augmentation.

\item We develop an adversarial training mechanism that ensures the synthetic classes generated through our mixing strategy maintain statistical authenticity by minimizing categorical distinctions between real and synthetic embeddings.

\item We achieve significant performance improvements in zero-shot inference, demonstrated through an 8\% improvement over baseline models in speaker verification tasks, with enhanced generalization to diverse speaker distributions and verified applicability across various zero-shot learning tasks.
%We demonstrate CAARMA's effectiveness on speaker verification tasks, showing improvements of 8\% over baseline models, and validate its broader applicability to other zero-shot speech analysis tasks.
\end{itemize}

%% file: sections/related-work.tex
\paragraph{Mixup.} The development of mixup strategies has evolved substantially since Mixup's original introduction by \citep{zhang2018mixup}, which generated virtual samples and mixed labels by linearly combining two input samples and their corresponding labels. This pioneering method proved particularly successful in enhancing data diversity and improving generalization in visual classification tasks. Extensions such as ManifoldMix \citep{verma2019manifold} applied this concept to hidden layers, while CutMix \citep{yun2019cutmix} introduced a patch-based approach by blending rectangular sections of images, offering a novel alternative for augmenting training data.
Subsequent mixup strategies focused on tailoring data mixing to specific contexts or improving the precision of mixing. Static policies like SmoothMix \citep{lee2020smoothmix}, GridMix \citep{baek2021gridmix}, and ResizeMix \citep{qin2023resizemix} used hand-crafted cutting techniques, while dynamic approaches such as PuzzleMix \citep{kim2020puzzlemix} and AlignMix \citep{venkataramanan2022alignmix} incorporated optimal-transport methods to determine mix regions with greater flexibility. For Vision Transformers, strategies such as TransMix \citep{chen2022transmix} and TokenMix \citep{liu2022tokenmix} focused on leveraging attention mechanisms to refine mixing operations, particularly for transformer architectures \citep{dosovitskiy2021an}.
Recent developments have adapted mixup techniques to tasks beyond classification, such as regression. C-Mixup, for instance, applies sample mixing based on label distances, using a symmetric Gaussian kernel to select samples that improve regression performance \citep{cai2021cmixup}. Further enhancing robustness, RC-Mixup integrates C-Mixup with multi-round robust training, creating a feedback loop where C-Mixup helps identify cleaner data, and robust training improves the quality of data for mixing \citep{liu2022rcmixup}. These specialized approaches reveal mixup's adaptability across various machine learning tasks, enhancing model performance and data resilience. However, it's important to note that none of these methods involve mixing in the embedding space, which could allow for the creation of entirely new and synthetic class identities,
\paragraph{Synthetic speech.}
Recent advancements in synthetic audio generation have emphasized the creation of diverse and high-quality datasets, crucial for training and evaluating audio-based AI models. A notable innovation is ConversaSynth, a framework utilizing large language models (LLMs) to generate synthetic conversational audio across varied persona settings \citep{gao2022conversasynth}. This method begins with generating text-based dialogues, which are then rendered into audio using text-to-speech (TTS) systems. The synthetic datasets produced are noted for their realism and topic variety, proving beneficial for tasks such as audio tagging, classification, and multi-speaker speech recognition. These capabilities make ConversaSynth a valuable tool for developing robust, adaptable AI models that can handle diverse audio data and complex conversational contexts. In the domain of speaker verification, SpeechMix introduces a novel method by mixing speech at the waveform level, carefully adjusting ratios to preserve the distinct characteristics of speaker identity \citep{jindal2020speechmix}. However, like many other generative and augmentation techniques, SpeechMix primarily focuses on manipulating known speaker voices rather than generating new identities, thereby limiting its utility for enhancing speaker diversity critical for effective zero-shot learning. In the context of speaker verification, discriminative neural clustering \cite{li2021discriminative} employs clustering techniques to enhance speaker diarization by learning discriminative embeddings, but it does not address class diversity through synthetic class generation as CAARMA does. Synthio employs a unique approach by using text-to-audio (T2A) diffusion models to augment small-scale audio classification datasets \citep{joassin2022synthio}. It enhances compositional diversity and maintains acoustic consistency by aligning T2A-generated synthetic samples with the original dataset using preference optimization. Additionally, exploring style transfer in synthetic audio, recent work by Ueda et al. employs a VITS-based voice conversion model, conditioned on the fundamental frequency (F0), to produce expressive variations from neutral speaker voices \citep{ueda2024expressive}. This method achieves cross-speaker style transfer in a FastPitch-based TTS system, incorporating a style encoder pre-trained on timbre-perturbed data to prevent speaker leakage. This technique enhances the utility of synthetic data in applications requiring rich stylistic diversity. These developments underline the increasing sophistication of synthetic audio generation techniques, from multi-speaker conversations in ConversaSynth to Synthio's optimized T2A augmentation for classification, and cross-speaker style transfers with VITS-based models. They collectively demonstrate the power of synthetic data to enrich audio model training by adding diversity and realism. However, these methods still face limitations in generating entirely new speaker identities, which is critical for expanding the range of recognizable voices in speaker verification systems. CAARMA addresses this gap by directly mixing in the embedding space, creating synthetic speakers that enhance zero-shot learning capabilities. CAARMA not only preserves speaker characteristics but also significantly expands the diversity of speaker identities, offering a superior solution for training more robust and adaptable speaker verification systems.
%\textbf{}

\begin{figure*}[ht]
\vskip 0.2in
\begin{center}
\centerline{\includegraphics[width=\textwidth]{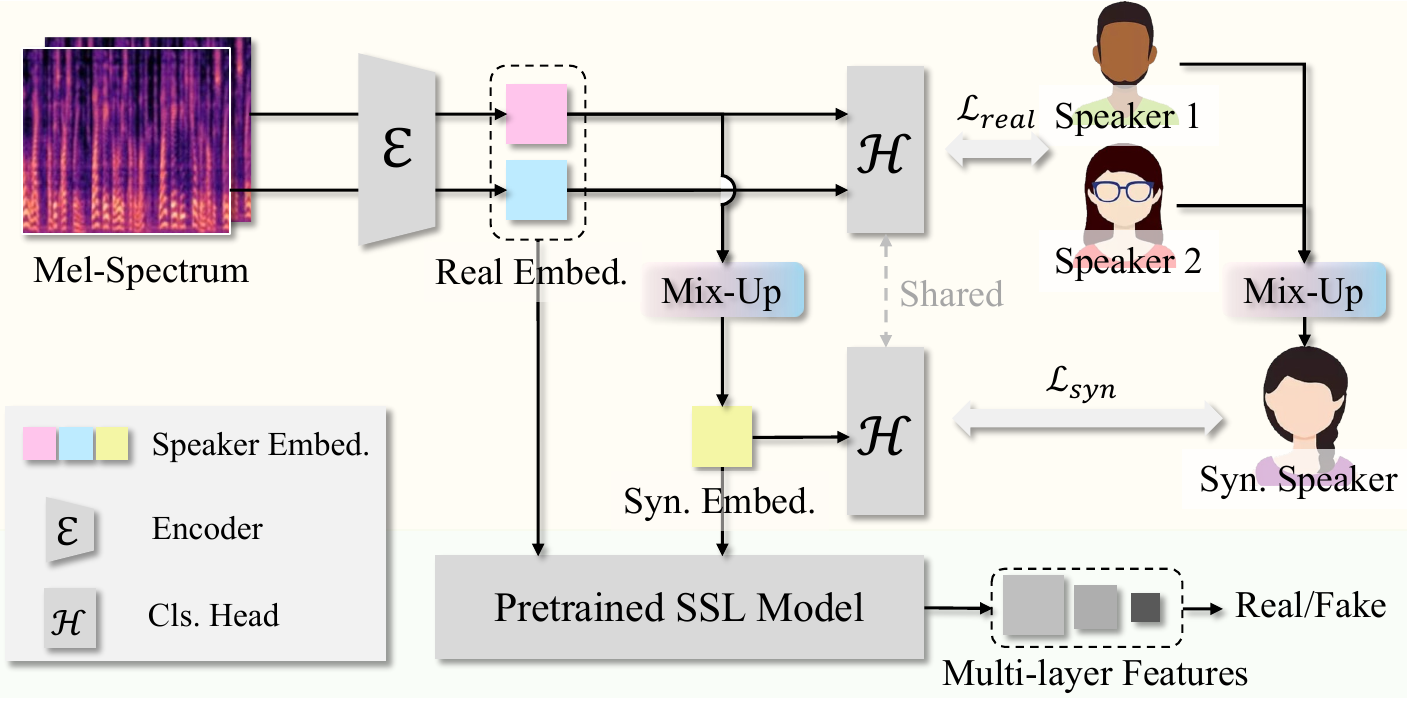}}
\caption{Illustration of CAARMA framework. (a) The encoder ($\mathcal{E}$) extracts embeddings from Mel-spectrograms, which are processed by a classification head ($\mathcal{H}$) for speaker identification and through Mix-Up for synthetic embedding generation. (b) Both real and synthetic embeddings are fed into a pretrained SSL model that acts as a discriminator, distinguishing between real and synthetic samples.}
%\caption{Illustration of the proposed framework. (a) The encoder ($\varepsilon$) generates embeddings ($e$), which are processed by SL-Mixup to create synthetic embeddings and synthetic labels. (b) Real ($R$) and synthetic ($S$) embeddings are fed into a self-supervised learning (SSL) model acting as a semantic discriminator, refining the embedding manifold through adversarial training. Note: Other loss components are omitted for clarity.}
\label{fig:pipeline}
\end{center}
% \vskip -0.2in
\end{figure*}

%% file: sections/method_2.tex
\subsection{Overview}
As mentioned in Section \ref{sec:intro}, ZSL models learn their ability to compactly cluster same-class embeddings while maintaining separation between classes primarily through exposure to a large number of classes during training; the more classes they are exposed to in training, the better they are able to generalize to unseen classes. In the speaker verification setting, this translates to training the model with recordings from a large number of speakers; the more the number of training speakers the better the model generalizes. To improve this generalization CAARMA attempts to increase the number of speakers by creating synthetic speakers while training.  Synthetic speakers may be created through generative methods such as \cite{cornell2024generating}; however this approach does not scale.  Instead, CAARMA creates them through a simple mixup strategy, as convex combinations of real speakers, with a key distinction: the mixup is performed in the \textit{embedding} space, where the classes are expected to form compact (and generally convex) clusters.  In order to ensure that these synthetic speakers are indeed representative of actual speakers, CAARMA utilizes a \textit{mixup discriminator}, a discriminator which attempts to distinguish between synthetic and real speakers: if this discriminator is fooled, the synthetic speakers are statistically indistinguishable from real ones.

When training the model, a conventional loss such as the Additive Margin Softmax (AM-Softmax) \cite{wang2018additive} is used. The synthetic classes, which are created dynamically during training, are included by dynamically  also expanding the class labels in the loss. In addition, the model also attempts to adversarially fool the discriminator. Once the model is trained, the discriminator is no longer needed and is discarded.

% We discuss the actual implementation below.

\subsection{Framework}
Our framework consists of three main components: an encoder for embedding generation, a synthetic label mixup mechanism for class augmentation, and an adversarial training scheme with a semantic discriminator. Figure~\ref{fig:pipeline} illustrates the complete pipeline of our approach.
The process begins with a waveform input that is transformed into a Mel-spectrogram. This spectrogram serves as input to the encoder $\mathcal{E}$, which generates embeddings $e$ that capture discriminative speaker characteristics. These embeddings undergo our SL-Mixup strategy, which generates synthetic embeddings $e_{\text{syn}}$ by mixing embeddings $e$ based on their closest neighbor weights $W$. Each synthetic embedding receives a corresponding synthetic label $ID_{\text{syn}}$ within the mini-batch.
The framework employs two primary loss functions: the encoder loss $\mathcal{L}_{\text{real}}$ for original embeddings and the synthetic loss $\mathcal{L}_{\text{syn}}$ for synthetic embeddings. A Self-Supervised Learning (SSL) model serves as the mixup discriminator to distinguish between real ($R$) and synthetic ($S$) embeddings. A discriminator loss $\mathcal{L}_{D}$ guides the discriminator to maximally distinguish between $R$ and $S$.
%, enhancing the learning of the embedding space. 
On the other hand, a generator loss $\mathcal{L}_{\text{gen}}$  guides the encoder to ``fool'' the discriminator, so that it perceives no distinction between real and synthetic embeddings.
\subsection{Encoder}
The encoder $\mathcal{E}$ transforms Mel-spectrograms into discriminative embeddings $e$ that capture speaker-specific acoustic features. We employ an MFA-Conformer model \cite{zhang2022mfa} as our encoder architecture, which combines feed-forward networks (FFNs), multihead self-attention (MHSA), and convolution modules. The model incorporates positional embeddings to handle variable-length input sequences effectively. For training, we utilize the AM-Softmax function %Additive Margin Softmax (AM-Softmax) function %\cite{wang2018additive}  
as our encoder loss $\mathcal{L}_{\text{real}}$.
%\ch{add loss equation} addressed
\[
\mathcal{L}_{\text{real}} = -\log \frac{e^{s \cdot (\cos(\theta_y) - m)}}{\sum_{j=1}^{C} e^{s \cdot \cos(\theta_j)}}
\]
where \( s \)  is a scaling factor used to stabilize gradients, \( C \) represents the number of classes, \( \cos(\theta_y) \) denotes the cosine similarity for the true class, and  \( m \) is an additive margin that enhances class separation by increasing inter-class distances.

\subsection{Synthetic Label Mixup}
Our SL-Mixup strategy generates synthetic embeddings $e_{\text{syn}}$ within each mini-batch by mixing embeddings $e$ according to their closest neighbor weights $W$, as detailed in Algorithm \ref{alg:mixup}. This approach ensures synthetic embeddings remain within the same manifold as real embeddings, avoiding arbitrary generation. The strategy creates synthetic labels $ID_{\text{syn}}$ and embeddings dynamically during training, enabling effective representation learning and facilitating the potential use of unlabeled data. To ensure that synthetic speakers are minimally confusable with their component speakers, we only combine pairs of speakers with a fixed weight of 0.5. This approach maintains a balanced contribution from each component speaker, preventing synthetic embeddings from collapsing into a single identity while maintaining inter-class separation. The synthetic loss $\mathcal{L}_{\text{syn}}$ is computed using the AM-Softmax loss function applied to synthetic embeddings $e_{\text{syn}}$. This loss is integrated into the main encoder loss $\mathcal{L}_{\text{real}}$, scaled by $1/\lambda$, where $\lambda$ represents the number of speakers. This integration ensures proper alignment of synthetic embeddings within the embedding manifold.
\begin{algorithm}[tb]
   \caption{SL-Mixup}
   \label{alg:mixup}
\begin{algorithmic}
   \STATE {\bfseries Input:} Feature matrix $X$, Label vector $Y$, Weight matrix $W$
   % \STATE {\bfseries Output:} Synthetic embeddings $X_{\text{syn}}$, Synthetic labels $Y_{\text{syn}}$, Weights for synthetic classes $W_{\text{syn}}$
   \STATE Initialize $W_{\text{syn}} \leftarrow \mathbf{0}$, $Y_{\text{syn}} \leftarrow \mathbf{0}$, $X_{\text{syn}} \leftarrow \mathbf{0}$
   \FOR{$y_i \in Y$}
      \STATE $\text{distances} \leftarrow \| \mathbf{W}[:, i] - \mathbf{W}[:, j] \|_2 \quad \forall j \in \text{label\_set} \setminus \{i\}$
      \STATE $\text{neighbor}(y_i) \leftarrow \arg\min(\text{distances})$
   \ENDFOR
   \FOR{$i \in \text{Batch}$}
      \STATE $l_1 \leftarrow Y[i], \; l_2 \leftarrow \text{neighbor}(l_1)$
      \STATE $W_{\text{syn}}[:,i] \leftarrow 0.5 \times (W[:,l_1] + W[:,l_2])$
      \STATE $Y_{\text{syn}}[i] \leftarrow \text{new\_label}(l_1, l_2)$
      \STATE $\text{index}[i] \leftarrow \text{find}(Y = l_2)$
      \STATE $X_{\text{syn}}[i] \leftarrow 0.5 \times (X[i] + X[\text{index}[i],:])$
   \ENDFOR
   \STATE {\bfseries Return:} $X_{\text{syn}}$, $Y_{\text{syn}}$, $W_{\text{syn}}$
\end{algorithmic}
\end{algorithm}
% change the notations here 
\begin{algorithm}[tb]
\caption{Adversarial Training with Synthetic Embeddings}
\label{alg:gan_training}
\begin{algorithmic}
\STATE {\bfseries Input:} Feature extractor $f(X)$, Model $M$, Discriminator $D$, Dataset $\mathcal{D}$ (waveforms $X$ and labels $Y$), Adversarial weight $\lambda_{\text{adv}}$
\FOR{each epoch $e \in [1, N_{\text{epochs}}]$}
\FOR{each batch $(X, Y) \in \mathcal{D}$}
\STATE Extract features $F = \text{Mel}(X)$
\STATE Compute embeddings $e = \mathcal{E}(F)$
\STATE Compute AM-Softmax loss $\mathcal{L}_{\text{real}}$
\STATE Generate synthetic embeddings $e_{\text{syn}}$ via mixup
\STATE Compute real predictions $D(e)$ and fake predictions $D(e_{\text{syn}})$
\STATE Compute discriminator loss:
\STATE $\mathcal{L}_{D} = \text{BCE}(D(e), 1) + \text{BCE}(D(e_{syn}), 0)$
\STATE Update $D$ using $\nabla \mathcal{L}_{D}$
\STATE Compute generator loss
\STATE $\mathcal{L}_{G} = \text{BCE}(D(e_{\text{syn}}), 1) + \text{BCE}(D(e), 0)$
\STATE Adjust $\lambda_{\text{adv}}$ based on $\mathcal{L}_{\text{real}} / \mathcal{L}_{G}$
\STATE Compute total loss $\mathcal{L}_{\text{total}} = \mathcal{L}_{\text{real}} + \lambda_{\text{adv}} \mathcal{L}_{G}$
\STATE Update $M$ using $\nabla \mathcal{L}_{\text{total}}$
\ENDFOR
\ENDFOR
\STATE {\bfseries Return:} Trained $M$ and $D$
\end{algorithmic}
\end{algorithm}

\subsection{Adversarial Training}
Our adversarial training process alternates between optimizing the encoder and discriminator, as described in Algorithm~\ref{alg:gan_training}. This optimization scheme continuously refines the embedding manifold through the interaction between real and synthetic embeddings:
\begin{itemize}
\item \textbf{Discriminator Training:} The discriminator $D$ learns to differentiate between real embeddings $e$ and synthetic embeddings $e_{\text{syn}}$ using features extracted from multiple model layers. The discriminator loss is defined as:
\begin{equation}
\mathcal{L}_{D} = \text{BCE}(D(e), 1) + \text{BCE}(D(e_{\text{syn}}), 0)
\end{equation}
where BCE represents binary cross-entropy loss.
\item \textbf{Generator Loss:} The encoder incorporates a generator loss $\mathcal{L}_{G}$ that guides embedding alignment with the manifold structure:
\begin{equation}
\mathcal{L}_{G} = \text{BCE}(D(e_{\text{syn}}), 1) + \text{BCE}(D(e), 0)
\end{equation}
\end{itemize}
\subsection{Mixup Discriminator}
To enhance the discriminative power of our framework, we incorporate a self-supervised model (HuBERT) \cite{hsu2021hubert} as a mixup discriminator. This discriminator leverages the pre-trained representations to provide richer gradients during adversarial training, improving the stability and quality of the learned embeddings.
The semantic discriminator processes embeddings through an adapter module that projects them into a compatible feature space. The adapter consists of a down-projection layer with spectral normalization, followed by fully connected layers with GELU activation \cite{hendrycks2016gaussian}. We employ skip connections and layer normalization to ensure stable training.
% \begin{equation}
% \resizebox{0.43\textwidth}{!}{%
% $\mathbf{h}_{\text{adapted}} = \text{Norm}2 \circ \text{Intermediate} \circ \text{Norm}1 (\mathbf{W}{\text{down}} \mathbf{x} + \mathbf{b}{\text{down}}) + \mathbf{x}$%
% }
% \end{equation}
The discriminator extracts features from multiple HuBERT \cite{hsu2021hubert} layers (7, 9, 11, and 12) to capture diverse speaker characteristics. These features are combined using learnable weights and processed through a residual classification block with spectral normalization and LeakyReLU activation \cite{xu2015empirical} to determine whether an embedding is real or synthetic. %This multi-layer approach ensures robust discrimination while maintaining computational efficiency.

%% file: sections/exp2.tex
\subsection{Datasets}
% testsets maybe training sets etc.. 
We utilize four datasets in our proposed approach: VoxCeleb1 \cite{nagrani2017voxceleb}, VoxCeleb2 \cite{chung2018voxceleb2}, and two datasets from the Dynamic-SUPERB \cite{huang2024dynamic} benchmark such as HowFarAreYou and DailyTalk datasets. The dataset statistics are summarized in Table \ref{tab:stats}. These datasets were employed across different tasks to evaluate the adaptability and generalizability of our pipeline:

\begin{itemize}

\item \textbf{Speaker Identification:} The primary task of our study involves speaker identification using VoxCeleb1 and VoxCeleb2. These large-scale datasets contain speech recordings from thousands of speakers. 
\item \textbf{Speaker Distance Estimation} The HowFarAreYou dataset originates from the 3DSpeaker dataset, designed to assess the distance of a speaker from the recording device. The task involves predicting distance labels (e.g., 0.4m, 2.0m) based on speech recordings. 
\item \textbf{Emotion Recognition:} We use the DailyTalk dataset to classify the emotional state (anger, disgust, fear, happiness, sadness, surprise, neutral) of a speaker based on speech utterances. This dataset contains speech samples labeled with seven distinct emotion categories. To maintain consistency with other datasets, we resample all recordings to 16 kHz before processing.
\end{itemize}

Table \ref{tab:stats} provides an overview of the datasets used in our experiments, including the number of classes and total utterances per dataset.

% The inclusion of these diverse datasets enables a comprehensive evaluation of our pipeline across multiple domains, including speaker recognition, spatial awareness, and emotional state classification. By testing on different types of tasks, we demonstrate the adaptability and robustness of our framework in learning meaningful representations across various speech-related challenges.

\begin{table}[t]
%\vskip 0.15in
\begin{center}
\begin{small}
\begin{sc}
\begin{tabular}{lcccc}
\hline
\textbf{ID} & \textbf{Dataset} & \textbf{Classes} & \textbf{Utterances} \\
\hline
1 & VoxCeleb1 & 1211 &  153,516 \\
2 & VoxCeleb2 &   5994 & 1,087135   \\
3 & HowFarAreyou &   3 &  3,000 \\
4 & DailyTalk & 7 &   16,600 \\
\hline
\end{tabular}
\caption{Dataset statistics used in our experiments.}
\label{tab:stats}
\end{sc}
\end{small}
\end{center}
\vskip -0.1in
\end{table}
\begin{table}[t]
%\vskip 0.15in
\begin{center}
\begin{small}
\begin{sc}
\begin{tabular}{lcccc}
\hline
ID & $L_{syn}$ & AT & MD & Results \\
\hline
1 & & & & 3.33 \\
2 & \checkmark & & & 3.28 \\
3 &  & \checkmark & & 3.15 \\
4 &  & \checkmark & \checkmark & 3.18 \\
5 & \checkmark & \checkmark & & 3.17 \\
6 & \checkmark & \checkmark & \checkmark & \textbf{3.09} \\

\hline
\end{tabular}
\caption{EER Results for MFA Conformer baseline, Adversarial Training (AT), Mixup Discriminator (MD), and Synthetic Loss $L_{syn}$ using VoxCeleb1 for the SV tasks.}
\label{tab:res_vox1_main}
\end{sc}
\end{small}
\end{center}
\vskip -0.1in
\end{table}

\begin{table}[t]
% \vskip 0.15in
\begin{center}
\begin{small}
\begin{sc}
\begin{tabular}{llll}
\hline
Encoder & Baseline & AT & AT+$\mathcal{L}_{syn}$ \\
\hline
ECAPA TDNN    & 4.22 & 3.96 & 3.87 \\
MFA Conformer & 3.33 & 3.18 & 3.09 \\
\hline
\end{tabular}
\caption{EER Results for two different encoders ECAPA-TDNN and MFA Conformer showing performance in baseline, Adeversarial Training (AT), and Synthetic Loss ($L_{syn}$) on VoxCeleb1-O.}
\label{tab:diff_enc_sv}

\end{sc}
\end{small}
\end{center}
\vskip -0.1in
\end{table}
\begin{table}[t]
%\vskip 0.15in
\begin{center}
\begin{small}
\begin{sc}
\begin{tabular}{lccc}
\hline
ID & Hidden Layers & EER (\%) & minDCF \\
\hline
1 & $h_{3},h_{6},h_{9},h_{12}$   & 3.22 & 0.31 \\
2 & $h_{6},h_{7},h_{8},h_{9}$    & 3.12 & 0.30 \\
3 & $h_{7},h_{9},h_{11},h_{12}$  & \textbf{3.09} & 0.28 \\
\hline
\end{tabular}
\caption{Ablation study of different hidden layers for Mixup Discriminator (MD) reporting EER (\%) and minDCF.}
\label{tab:hidden_res}
\end{sc}
\end{small}
\end{center}
\vskip -0.1in
\end{table}
\begin{table*}[tb]
    \centering
    \renewcommand{\arraystretch}{1.2} % Adjust row spacing
    \begin{tabular}{l c c c c c c}
        \hline
        \textbf{ID} & \textbf{Model} & \textbf{\# Parameters} & \textbf{EER(\%)} & \textbf{minDCF} & \textbf{Utterances} & \textbf{\# Speakers} \\
        \hline
        1 & MFA-Conformer& 19.8M & 12.82 & 0.770 & 12,335 & 100 \\
        2 & MFA-Adversarial & 19.8M & 11.83 & 0.790 & 12,335 & 100 \\
        \hline
        3 & MFA-Conformer& 19.8M & 0.86 & 0.066 & 1,240,651 & 7,205\\
        4 & MFA-Adversarial & 19.8M & 0.81 & 0.036 & 1,240,651 & 7,205 \\

        \hline
    \end{tabular}
    \caption{\textit{Performance overview of all systems on VoxCeleb1-O}}
    \label{tab:sv_big_res}
\end{table*}

\subsection{Experimental Setup}

\subsubsection{Model Architecture}
We train two baseline architectures for speaker verification:

\textbf{ECAPA-TDNN} \cite{desplanques2020ecapa}: Contains three SE-Res2Blocks with 1024 channels (20.8M parameters).

\textbf{MFA-Conformer} \cite{zhang2022mfa}: Employs 6 Conformer blocks with 256-dimensional encoders, 4 attention heads, and convolution kernel size of 15 (19.7M-20.5M parameters).

Both architectures generate 192-dimensional embeddings for fair comparison. 
For emotion and distance tasks, we utilize HuBERT-Large (pretrained on LibriSpeech) with 1024-dimensional embeddings and 768 hidden units.

\subsubsection{Adversarial Training}
We incorporate adversarial training into our baseline experiments. In this approach, each model is retrained from scratch within our adversarial framework. The discriminator is trained concurrently with the encoder. The discriminator's role is to effectively distinguish between real and synthetic embeddings, enforcing a well-structured representation. 

\paragraph{Mixup discriminator} To determine the most informative HuBERT hidden layers for speaker representation, we conduct an ablation study. We experiment with different layer configurations, including \(\{h_{3}, h_{6}, h_{9}, h_{12}\}\), \(\{h_{7}, h_{9}, h_{11}, h_{12}\}\), \(\{h_{6}, h_{7}, h_{8}, h_{9}\}\) and evaluate their impact on model performance.

%uses selected hidden layers to extract speaker-specific features. This discriminator learns to differentiate between real and synthetic embeddings, guiding the encoder to learn a more structured embedding space. We conduct 3 different experiments with different layer configurations and evaluate impact on model performance.

\subsubsection{Implementation Details}
We implement all baseline systems and discriminators using the PyTorch framework \citep{yun2019cutmix}. Each utterance is randomly segmented into fixed 3-second chunks, with 80-dimensional Fbanks as input features, computed using a 25 ms window length and a 10 ms frame shift, without applying voice activity detection. All models are trained using AM-Softmax loss with a margin of 0.2 and a scaling factor of 30. We use the AdamW optimizer with an initial learning rate of 0.001 for model training, while the discriminator is optimized separately with AdamW at an initial learning rate of 2e-4. To prevent overfitting, we apply a weight decay of 1e-7 and use a linear warmup for the first 2k steps, though no warmup is applied to the discriminator. Training is conducted on NVIDIA V100 GPUs with a batch size of 50, and all models are trained for 30 epochs.

\paragraph{Computational Complexity:}CAARMA introduces no additional computational cost during inference, as the discriminator is not used, and inference remains identical to the baseline model. During training, the primary computational overhead arises from two components: (1) the forward and backward passes through the discriminator, and (2) the computation of the AM-Softmax loss for both real and synthetic (mixup) data. Since synthetic embeddings are dynamically generated from real-data embeddings, no additional embedding computations are required. In our experiments, synthetic data is generated in a 1:1 ratio with real data, effectively quadrupling the compute time for the loss calculations (doubling for AM-Softmax on real and synthetic data, and doubling for the discriminator). However, as the loss computation constitutes a minor fraction of the overall training cost, the total increase in training time remains modest. Additionally, since mixup data is computed dynamically, the memory overhead is negligible.
% Encoder speakers ecapa mfa hubert mention numbers and training etc 
% discriminator and dim and lr schedulers. .

%% file: sections/results.tex
\begin{table}[t]
\begin{center}
\begin{small}
\begin{sc}
\begin{tabular}{lll}
\hline
Data set & Baseline & AT \\
\hline
Emotion Classification    & 83\% & 85.50\% \\
HowFarSpk & 77.91\% & 79.97\%\\
\hline
\end{tabular}
\caption{Classification accuracies for Hubert Encoder baseline and with Adversarial Training on two different tasks.}
\label{tab:diff_tasks}
\end{sc}
\end{small}
\end{center}
\end{table}

To validate the effectiveness of our approach, we conduct comprehensive experiments across speaker verification, emotion classification, and speaker distance estimation tasks. Our analysis demonstrates significant improvements through adversarial refinement on model generalization.

\subsection{Speaker Verification Task}
We evaluate our models on VoxCeleb1-O, the official test set of VoxCeleb1. For evaluating the performance, we use the Equal Error Rate (EER) and minimum Detection Cost Function (minDCF). EER represents the point where the false acceptance rate equals the false rejection rate, providing a single measure of verification accuracy (lower is better). The minDCF quantifies the cost of detection errors, balancing false positives and false negatives, with lower values indicating better performance. These metrics assess the model's ability to distinguish between same-speaker and different-speaker pairs effectively.
\subsubsection{Small Scale}
Our initial evaluations focused on models trained on VoxCeleb1 to facilitate thorough experimentation with different model configurations.

Using the MFA Conformer as the Encoder, we conduct several experiments (Table \ref{tab:res_vox1_main}) reporting the EER. The addition of synthetic loss alone (Model $ID_{2}$) yield slight improvements over the baseline (Model $ID_{1}$). More substantial gains were achieved through adversarial training (Model $ID_{3}$), with further improvements observed when incorporating the mixup discriminator (Model $ID_{4}$). The best performance was achieved by Model $ID_{6}$, which combined all three components: synthetic loss, mixup discriminator, and adversarial training.

To further validate the generalizability of our approach, we implement it with an alternative speaker encoder. As shown in Table \ref{tab:diff_enc_sv}, the combination of adversarial training and mixup discriminator improved performance by 6.56\% compared to the baseline. Adding synthetic loss further enhanced the improvement to 8.29\%.

We conduct an ablation study to identify the most informative hidden layers for speaker representation. Table \ref{tab:hidden_res} presents the EER and minDCF across various layer configurations. Our analysis revealed that layers 7, 9, 11, and 12 provide the most effective speaker characteristics representation, suggesting that later layers capture more valuable speaker-specific information.

\subsubsection{Large Scale}
To demonstrate scalability, we train on the combined VoxCeleb1 and VoxCeleb2 datasets, creating a substantially larger training corpus. As summarized in Table \ref{tab:sv_big_res}, the MFA-Conformer encoder (Model $ID_{3}$) achieves strong baseline performance, which is further enhanced through adversarial refinement (Model $ID_{4}$).  

Specifically, adversarial training reduces the EER from 0.86\% to 0.81\% and nearly halves the minDCF from 0.066 to 0.036. These improvements are consistent with the small-scale trends, confirming that our framework scales robustly to larger and more diverse speaker populations.

Importantly, we note that even in the limited-diversity setting of VoxCeleb1—where only 100 speakers are present—our approach still yields consistent gains (Model IDs~1–2). This suggests that the framework is not merely leveraging broader class coverage in large-scale datasets, but is also effective in scenarios with constrained speaker diversity. In practice, this means that our method improves speaker verification robustness both under resource-rich conditions and under more restrictive data availability.  

Overall, these results highlight that adversarial training is beneficial across training regimes: it generalizes well to large-scale, diverse datasets while still offering tangible improvements when diversity is limited.

\subsection{Emotion and Speaker Distance Tasks}

To demonstrate the generalizability of our approach across different speech processing domains, we evaluate its effectiveness on emotion classification and speaker distance estimation using the DailyTalk and HowFarAreYou test sets, respectively. As shown in Table \ref{tab:diff_tasks}, our method improved classification accuracy across both tasks: emotion classification accuracy increased from 83\% to 85.50\%, while speaker distance estimation improved from 77.91\% to 79.97\%. These results demonstrate that our approach can be effectively integrated with various models across various domains.

% \begin{table}[t]
% \caption{EER Results for different discriminators along with MFA Conformer Encoder on the SV tasks.}
% \label{sample-table}
% \vskip 0.15in
% \begin{center}
% \begin{small}
% \begin{sc}
% \begin{tabular}{lcc}
% \toprule
% Discriminator & GAN & GAN+$Loss_{syn}$ \\
% \midrule
% Hubert    & 3.18 & 3.09 \\
% PatchGAN & 3.19 & 3.15 \\
% \bottomrule
% \end{tabular}
% \end{sc}
% \end{small}
% \end{center}
% \vskip -0.1in
% \end{table}

%% file: sections/conclusion.tex
In this work, we introduce CAARMA, a novel class augmentation framework designed to tackle the challenge of limited class diversity in zero-shot inference tasks. Our approach synthesizes strategic data mixing with an adversarial refinement mechanism to align real and synthetic classes effectively within the embedding space. 
We validate CAARMA's effectiveness in speaker verification, achieving an 8\% improvement over baseline models, and extend its application to emotion classification and speaker distance estimation, where it also shows significant gains. These results underscore CAARMA’s capability to enhance embedding structures in various zero-shot inference scenarios. Our framework offers a scalable solution to the class diversity problem, facilitating integration into existing systems without the need for new real-world data collection. With our code released publicly, we anticipate that CAARMA will aid both research and practical applications in zero-shot learning. Future work will focus on expanding CAARMA’s utility to larger datasets and other domains, such as computer vision.